# DAΦNE

S. Guiducci, D. Alesini, G. Benedetti, M.E. Biagini, C. Biscari, R. Boni, M. Boscolo, A. Clozza, G. Delle Monache, G. Di Pirro, A. Drago, A. Gallo, A. Ghigo, F. Marcellini, G. Mazzitelli, C. Milardi, L. Pellegrino, M.A. Preger, P. Raimondi, R. Ricci, C. Sanelli, M. Serio, F. Sgamma, A. Stecchi, C. Vaccarezza, M. Zobov, LNF, Frascati, Italy

*Abstract*

The results of 2002 DAΦNE operation for the two experiments KLOE and DEAR are described. During 2003 a long shutdown has been dedicated to the installation of new Interaction Regions (IR) and to hardware modifications and upgrades. In the last section optics studies and performances expectations for the new machine configuration are reported.

## KLOE AND DEAR RUNS

Peak, daily and integrated luminosity delivered to KLOE and DEAR experiments in 2002 is shown in Fig.1.

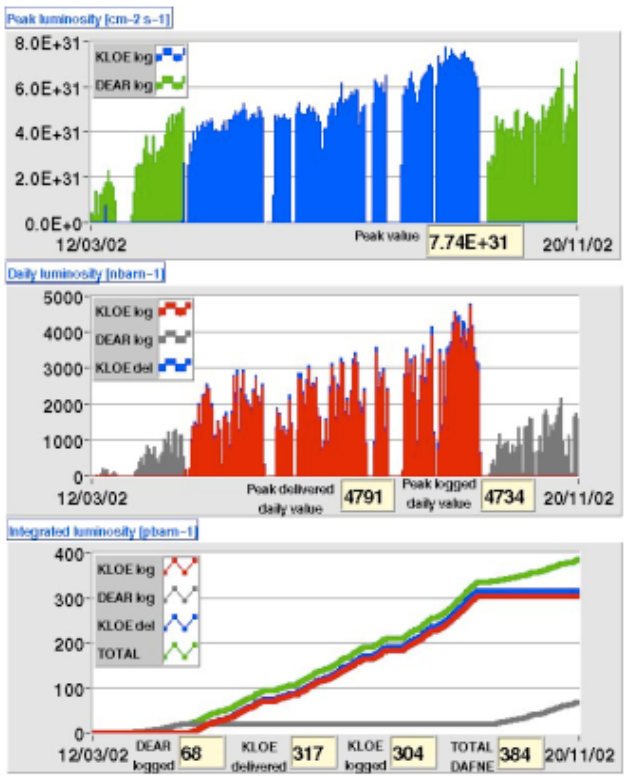

Fig.1- Peak, daily and integrated luminosity for 2002.

During 2002 four months were dedicated to DEAR operation. This allowed to optimize the machine configuration obtaining nearly the same luminosity as for KLOE and to complete the first phase of the DEAR physics program. The results of machine performance during 2002 runs are listed in Table 1.

KLOE IR has permanent magnet low-β quadrupoles while DEAR has electromagnetic ones. This makes a difference between the two machine configurations: when the beams are colliding in KLOE, it is possible to remove the low β and keep the beams well separated at the other Interaction Point (IP), thus improving the luminosity. This is not possible for DEAR. The main machine parameters in the two configurations are listed in Table 2.

Table 1 – 2002 machine performance

| Configuration | KLOE | DEAR |
|---|---|---|
| Number of bunches | 49 + 49 | 95+95 |
| Total current per beam (A) | 0.8/1.1 | 1.3/1.0 |
| Peak luminosity (cm$^{-2}$s$^{-1}$) | .75 x10$^{32}$ | .7 x10$^{32}$ |
| Beam-beam tune shift | ~ .02 | ~ .016 |
| Average luminosity (cm$^{-2}$s$^{-1}$) | 0.5 x10$^{32}$ | 0.2 x10$^{32}$ |
| Best daily luminosity (pb$^{-1}$) | 4.8 | 2.2 |
| Luminosity lifetime (h) |  | 0.6 |
| Number of fillings per hour | ≈ 3 | ≈ 1.7 |
| Injection frequency e-/e+ (Hz) | 2/1 | 2/1 |
| Data acquisition during inject. | on | off |

Table 2 – Main machine parameters

|  | KLOE | DEAR |
|---|---|---|
| ε ( m·rad) | .74 10$^{-6}$ | .62 10$^{-6}$ |
| $\beta_x$ ( m) @ IP$_{KLOE}$ | **2.7** | 1.7 |
| $\beta_y$ ( m) @ IP$_{KLOE}$ | **.026** | 13. |
| $\beta_x$ ( m) @ IP$_{DEAR}$ | 2.7 | **1.7** |
| $\beta_y$ ( m) @ IP$_{DEAR}$ | .09 | **.030** |
| θ( mrad) | ±11.5 | ±14.5 |

During 2002 the integrated luminosity delivered to KLOE experiment (300 pb$^{-1}$ in 5 months) was larger by a factor 1.5 with respect to 2001 (200 pb$^{-1}$ in 7 months).

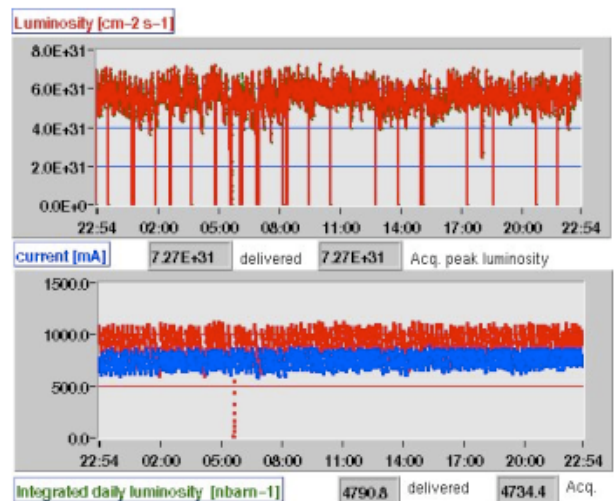

Fig. 2 – KLOE currents and luminosity in 24 hours

The βs at the IP were lowered ($\beta_y$ from .03 to .026 m, $\beta_x$ from 5.6 to 2.7 m) to increase the luminosity. The reduction of the horizontal β at the IP was also very effective in reducing machine background in the detector. Moreover a continuous optimization of the parameters affecting background, lifetime, and luminosity was performed: old and new scrapers' position, closed orbit correction, crossing angle, separation in the non colliding IR, working point in the tune plane, coupling, dynamic aperture (by tuning sextupole and octupole strengths and βs at sextupoles and wigglers), horizontal emittance (0.96μm ⇨ 0.76μm)

The peak luminosity was increased by a factor 1.5 with respect to 2001 [1] and the machine background in the detector was reduced by a factor ~3. Luminosity and currents for a typical KLOE run (24 hours) are shown in Fig. 2. The average luminosity is kept very close to the peak one by injecting with data acquisition on.

## 95 BUNCHES OPERATION FOR DEAR

During DEAR runs DAΦNE has been operated by filling 95 consecutive buckets out of 120, with an ion clearing gap [2]. This operation was made possible by an optics modification and by a very accurate setup of transverse and longitudinal feedbacks.

### Optics modification

To reduce the effect of parasitic crossings, which occur at 0.4 m distance, the optics of the DEAR IR has been modified by lowering the horizontal β-function at the IP and by increasing the crossing angle.

The FDF low beta triplet, optimized for 10 mrad crossing angle, was changed to a DF doublet which yields a larger crossing angle and a smaller value of the β functions at the IP, with the same chromaticity.

With a crossing angle θ = ±14.5 mrad and $\beta_x$ = 1.7 m the distance between the beam centres at the first parasitic crossing is 11 $\sigma_x$ and therefore the beam-beam effect on the particles in the tails of the distribution does not affect beam lifetime.

### Longitudinal quadrupole instability

During 2002 this phenomenon has been deeply investigated [3]. It appears in both rings, at high currents, but with different single bunch thresholds: ~20% higher for $e^+$ than for $e^-$. The positron beam power spectrum with 100 bunches, 900mA, during collision is shown in Fig. 3. It exhibits a large longitudinal quadrupole instability.

The longitudinal feedback is able to control the instability. A multi-mode filter response has been implemented to simultaneously optimize the longitudinal feedback performances at the zero, dipole and quadrupole modes. This allowed storing stable beams at high currents (more than 1.8 A) in 95 contiguous bunches. An increase of peak luminosity due to the smaller $\beta_y^*$ value was achieved in April: L = 0.45 $10^{32}$ cm$^{-2}$ s$^{-1}$ with 47 bunches.

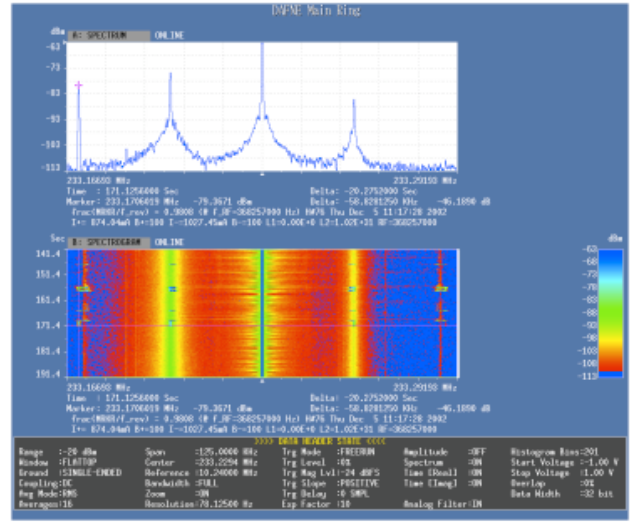

Fig. 3 - Positron beam power spectrum with 100 bunches, 900mA, during collision

In November, doubling the number of bunches, a peak luminosity of 0.7 $10^{32}$ cm$^{-2}$ s$^{-1}$, at the same background level, was obtained. The results of 95 bunches operation are shown in Table 1.

## MAIN HARDWARE MODIFICATIONS

In 2003 a long shutdown was dedicated to the following hardware installations:
    Finuda IR
    New Kloe I.R.
Moreover various systems were modified:
    Injection straight sections and kickers
    Scrapers
    Bellows
    Ion Clearing Electrodes
    Wigglers.

### New Interaction Regions

The simple DEAR IR (electromagnetic quadrupoles, no solenoid, very small detector) was replaced by the FINUDA one with a large detector, high field solenoid (B = 1 T), compensating solenoids and permanent magnet quadrupoles inside the detector. The supports allow variable quadrupole rotation to compensate the coupling at different magnetic fields (from 0 to maximum) in the solenoids.

KLOE IR was taken out from the detector, removing all the elements in the nearby straight sections, modified and inserted again. The supports were modified as in FINUDA to provide variable quadrupole rotation. In order to decrease the IP beta-functions and the lattice chromaticity the FDF triplet was converted into a DF doublet by removing the first F quadrupole and increasing the gradient integral of the other one by adding a new permanent magnet quadrupole adjacent to it. Masks to optimise background rejection were inserted.

These new IRs will allow 100 bunches operation because, as demonstrated in last DEAR shifts, $\beta_x$ at the IP is low enough to make the parasitic crossing not critical.

*Injection straight sections*

Injection straight sections have been rearranged, adding two quadrupoles and one sextupole, to improve injection efficiency and dynamic aperture. The kicker pulse length has been shortened in order to improve multibunch injection efficiency, at the price of a smaller peak field; this has been compensated by increasing the $R_{12}$ element of the transfer matrix between the kicker and the injection septum.

*Scrapers, bellows and Ion Clearing Electrodes*

The overall scrapers efficiency is satisfactory but some had problems with the tapers. Tapers were removed in the horizontal ones, less critical to the ring impedance, and modified in the vertical ones.

Some copper bellows were found to be distorted; they have been flattened with the insertion of pins.

About 50% of Ion Clearing Electrodes were broken due to faulty welding. Most of them were replaced with welding-free electrodes.

*Wiggler field modifications*

The strong sextupole components and the field roll off at large offsets in the wiggler magnetic field produced a significant reduction of the dynamic aperture [4,5]. A new wiggler, identical to those installed on the collider, was purchased from the same builder, in order to find a proper pole shape capable of reducing the nonlinearities. The resulting profile [6] was applied to a set of iron plates glued on the poles of all wigglers in the rings. The horizontal distribution of the field at pole center is shown in Fig. 5 for the different pole profiles tried during the optimization. Comparison between the field with new plates, not modified, and the final configuration shows a larger extension of the flat field region.

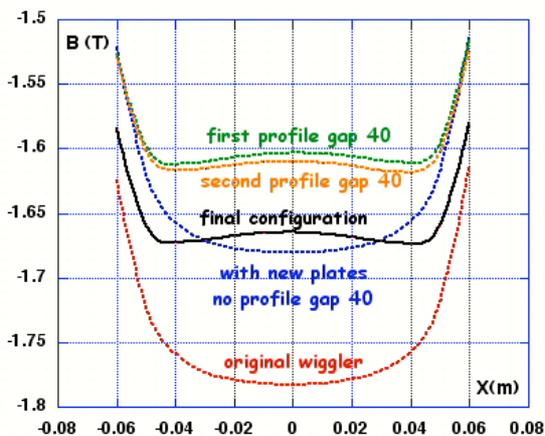

Fig. 5 - horizontal field distribution at pole center for different configurations. The solid line is the final one.

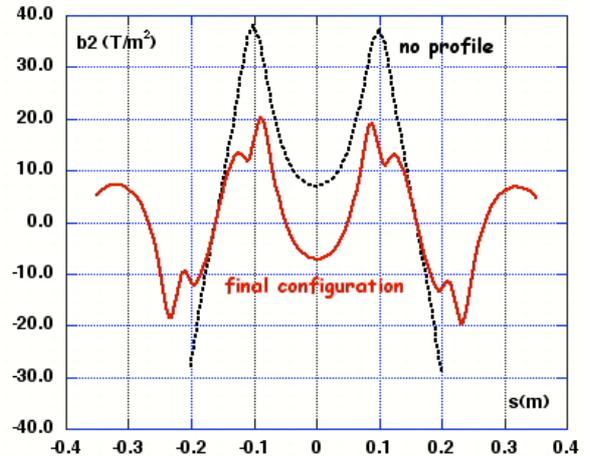

fig. 6 - Sextupole component on the beam axis for the original (dashed) and modified (solid) wiggler pole profile.

Moreover the sextupole component on the beam axis, shown in Fig. 6, has been significantly reduced in the final configuration so that its average over the polealmost vanishes.

One of the two terminal poles has been modified in order to increase its sextupole component. This modification is beneficial to the dynamic aperture, at least as the reduction of the main poles non-linearity.

An example of the improvement of the dynamic aperture is given in Fig. 7. Particle trajectories with initial value x = (2, 4, 6, 8 etc.) $\sigma_x$ are shown.

## OPTICS STUDIES

A new lattice for FINUDA, which will be used at machine start-up, has been designed. The main characteristics are summarized below:

    Low emittance: 0.42μm
    Low betas in the wigglers to minimize effect of non-linearity
    IP $\beta_x$=1.7m to allow 100 bunches operation
    IP $\beta_y$=27mm, nearly equal to the bunch length
    Additional sextupoles in the wigglers and at the septum
    Optimized phase advance between the sextupoles
    Low invariants to minimize background
    Straight sections optimized for injection efficiency and dynamic aperture

FINUDA will take data in the first months of 2004, then will be moved out of the beam line and the machine will be operated for KLOE.

The KLOE lattice, based on the new IR, will have the same characteristics listed above, moreover the low-β in the second IR will be removed. Therefore a large beam separation and a lower chromaticity will allow a higher luminosity. During 2004 we expect to reach in 100 bunches operation the same single bunch luminosity obtained for KLOE in 50 bunches.

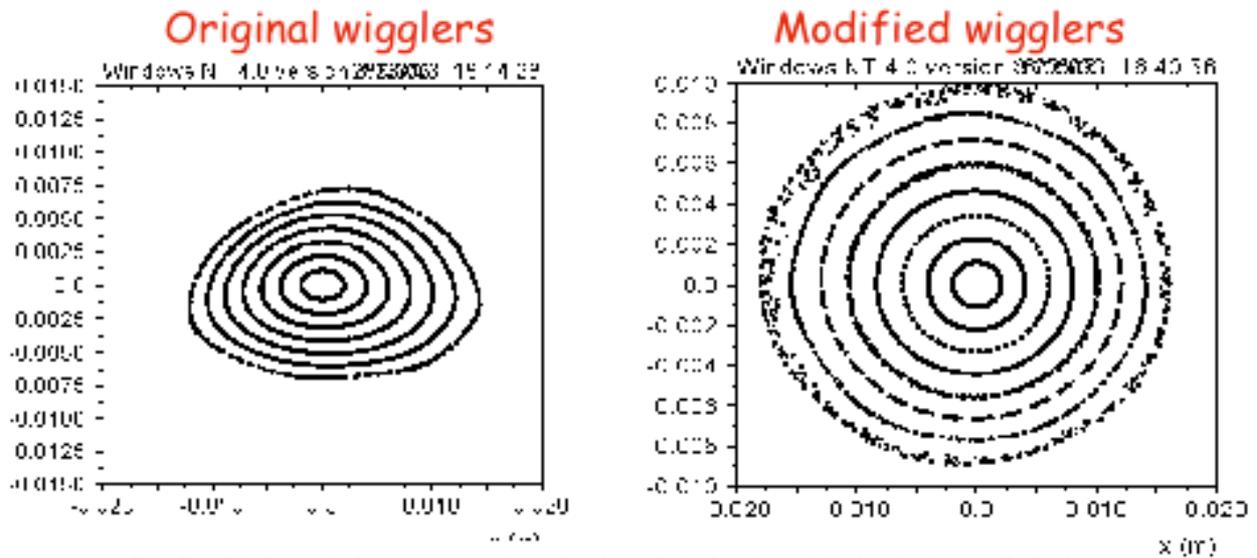

Fig. 7 – Horizontal phase space trajectories for original and modified wigglers at Δp/p=0.

The performance expected for 2004 runs, extrapolated by 2002 results and taking into account the improvements made during the shutdown, is summarized in Table 3.

Table 3 - Expected performance for 2004 runs

| Number of bunches | 100-110 |
|---|---|
| Current/beam | 2 A |
| Single bunch luminosity | $2\ 10^{30}$ cm$^{-2}$ s$^{-1}$ |
| Lifetimes | >1hr |
| Daily integrated luminosity | 10 pb$^{-1}$ |
| Integrated luminosity | 2.0 fb$^{-1}$/year |

Further studies to increase the luminosity are in progress. The reduction of machine nonlinearity should allow more flexibility in choosing the working point in the tune plane. Therefore it should be possible to minimize the beam-beam blow-up by exploring different working points much closer to the integer or half-integer (as in other factories).

The possibility to reduce the bunch length by using a negative momentum compaction has been discussed. Measurements done at KEK confirm this effect. Tests of a negative momentum compaction lattice are foreseen in DAΦNE to study if it can be used for collision in order to further reduce the vertical beta function at the IP and increase luminosity. [7,8].

## PRESENT STATUS

Hardware installation has been completed mid July and all machine subsystems have been checked. Commissioning of the rings with the new FINUDA optics just started, beams have been already stored in both rings.